\documentclass[twocolumn,aps,prb,showpacs]{revtex4-1}
\usepackage[utf8]{inputenc}
\usepackage[normalem]{ulem}
\usepackage[T1]{fontenc}
\usepackage[unicode=true, pdfusetitle, bookmarks=true, bookmarksnumbered=false, bookmarksopen=false, breaklinks=false, pdfborder={0 0 1}, backref=false, colorlinks=false]{hyperref}
\hypersetup{colorlinks,linkcolor=blue,citecolor=blue,urlcolor=blue}
\usepackage{float}
\usepackage{verbatim}
\usepackage[abs]{overpic}
\usepackage{color}
\usepackage{comment}
\usepackage{amsmath,amssymb}
\usepackage{graphicx}
\usepackage{subcaption}

\begin{document}
\title{The intermediate fermionic species created by $SO(3)$ rotation in the representation of the Dirac equation}
\author{H. Moaiery}
\affiliation{Jundi-Shapur University of Technology, Desful, Iran}

\author{M. N. Najafi}
\affiliation{Department of Physics, University of Mohaghegh Ardabili, P.O. Box 179, Ardabil, Iran}

\begin{abstract}
The question of how does the Dirac equation depend on the choice of the $\gamma$ matrices has partially been addressed and explored in the literature. In this paper we focus on this question by considering a general form of $\gamma$ matrices, and call the resulting spin $\frac{1}{2}$ fermions as \textit{intermediate fermion species} (IFS). By inspecting the properties of IFS, we find that all species transform to each other by a $SO(3)$ similarity transformation in the space of parameters, that are the entities of the $\gamma$ matrices. Many properties, like eigenvalue problem and boost are tested for IFS. We find also sub-representations that generate Majorana fermions, which is isomorphism to $U(1)$ group.
\end{abstract}


\maketitle

\section{Introduction}

The representations of fermions governed by the Dirac equation have vast applications in various fields in the fundamental and theoretical physics, ranging from elementary particles~\cite{gasiorowicz1966elementary} and quantum chromodynamics~\cite{ioffe2010quantum} to condensed matter~\cite{morandi2013field}, photonics ~\cite{sansoni2012two}, and superconductivity~\cite{chamon2010quantizing}. Three important representations of the Dirac equation are the Dirac fermions, the Weyl fermions and the Majorana fermions~\cite{leijnse2012introduction}, depending on the choice of the matrices in the Dirac equation (namely the $\gamma$ matrices), which show different properties in some aspects~\cite{ryder1996quantum}. The examples in condensed matter physics are intrinsic graphene for which the electrons show linear (gapless) dispersion realizing the massless Dirac fermions~\cite{neto2009electronic,sarma2011electronic,peres2010colloquium}, and the edge states in quantum Hall effect which are Majorana fermions~\cite{nayak2008non}. Weyl fermions may be realized as emergent quasiparticles in a low-energy condensed matter system. In Weyl semimetals~\cite{xu2015discovery,xu2015discovery2,xu2015experimental,armitage2018weyl} (as a topologically nontrivial phase of matter) the low energy excitations are Weyl fermions that carry electrical charge, with distinct chiralities.
Symmetry-protected Dirac fermions in topological insulators~\cite{moore2010birth,hasan2010three,kane2005z,gu2009tensor,pollmann2012symmetry,chen2011classification}, Majorana fermions in low energy excitations of condensed matter systems~\cite{elliott2015colloquium,beenakker2015random}, like quantum Hall effect~\cite{nayak2008non} and superconductivity~\cite{elliott2015colloquium}, are the other examples of emergent fermions with vast applications. Not surprisingly, the representation of fermions in the standard model of particles also plays a dominant role~\cite{avignone2008double,kuno2001muon}\\

Despite of this huge applications which stimulated a huge rate of studies on emergent fermionic systems (putting the subject in a large class ``Dirac material''~\cite{chiu2016classification}), there is no comprehensive systematic investigation on the possible \textit{intermediate fermionic species (IFS)} which arises by inspecting the possible representations of the Dirac equation. To be more precise, only a few specific and useful forms of fermions have been was considered, such as the standard or the Dirac-Pauli Representation (shown here by an S-index), supersymmetric representation, Weyl or spinor representation, Majorana representation etc., corresponding to appropriate choice of $\gamma$ matrices~\cite{ryder1996quantum}. Here we show that the Dirac equation admits the existence of IFSs by a systematic continuous $SO(3)$ rotation in the representation space of the Dirac equation, i.e. introducing and analyzing the general form of the $\gamma$ matrices. We argue that, although the ``type'' of the fermions change by this rotation, some properties of the particles like the dispersion relation and the Klein tunneling do not change, leading us to consider it as a symmetry of the Dirac equation. Many properties of the IFS particles are investigated in terms of rotation parameters, like the helicity and boost properties, parity. We also investigate the intermediate Majorana species as a sub-representation of $SO(3)$ which is homomorphic to $U(1)$ group.\\

The paper has been organized as follows: In the next section we introduce the IFS particles as the solution of the Dirac equation with general $\gamma$ matrices. Besides finding the wave functions, we investigate the behavior of the IFS under boost transformation. In SEC.~\ref{SEC:SO3} we introduce $SO(3)$ rotations which transform the IFS particles. Section~\ref{SEC:subr} is devoted to sub-representations of the $SO(3)$ group, i.e. the $U(1)$ group. In SEC.~\ref{SEC:Trans} we explain how to find IFS particles from the Dirac fermions. We close the paper by a conclusion.

\section{General representation of linearized relativistic particles}

In the natural units $\hbar=c=1$ the Dirac equation in $d+1$ dimensions is
\begin{equation}
\left( \gamma^{\mu}p_{\mu}\mp m\right) \Psi\left( \vec{x},t\right) =0
\label{Eq:DiracEq}
\end{equation}
where $p^{\mu}$ is four- [or generally $(d+1)$-] momentum ($\mu=0,1,...,d$, in which $0$ stands for the time component, and the others for the spatial components), and $ \gamma^{\mu} $s are the (even dimensional) Dirac matrices that are not unique (the possible minimum dimension of which depends on $d$)~\cite{ryder1996quantum}. As a well-known fact, the above equation gives rise to the Klein-Gordon equation, imposing some strong limitations on the choice (representation) of the $\gamma$ matrices. Having chosen the representation of the $\gamma$ matrices, one may reach to the other representations by a simple similarity transformation. To be more precise, let us suppose that we have 
\begin{equation}\label{Eq:transformation-gamma}
\gamma^{\mu}=T\gamma^{\mu}_DT^{-1}
\end{equation}
where $\gamma^{\mu}_D$ are the Dirac gamma matrices, and $T$ is a general transformation. Then obviously the same Dirac equation is valid, i.e. $\left( \gamma^{\mu}_Dp_{\mu}\mp m\right) \Psi_D\left( \vec{x},t\right) =0$, where $\Psi\equiv T\Psi_D$ is the solution of the Dirac equation in the Dirac representation. It is the aim of the present paper to study systematically this problem by considering an arbitrary form of $\gamma$ matrices and find the possible forms that $T$'s can have. We argue about some possible non-trivial outcomes and consequences of this ``generalization'', postponing any further investigations and uncovering any possible consequences to the community and also our future works.\\

\subsection{General $\gamma$s and wave functions}
Let us start with the general expectation that the Klein-Gordon equation casts to a product of two copies of Eq.~\ref{Eq:DiracEq} with the requirement~\cite{ryder1996quantum}
\begin{equation}
\gamma^{\nu}\gamma^{\mu}p_{\nu}p_{\mu}=g_{\text{M}}^{\mu\nu}p_{\nu}p_{\mu}=m^{2}, 
\end{equation} 
where $ g_{\mu\nu}^{\text{M}}=\text{diag}\left(+1,-1,\cdots,-1 \right) $ is the symmetric Minkowski metric, giving rise to
\begin{equation}
\left\lbrace \gamma^{\mu},\gamma^{\nu}\right\rbrace=2g^{\mu\nu}I,
\end{equation}
where $I$ is the $d+1$ dimensional identity matrix. Throughout this paper we use the following Hermitian matrices:
\begin{equation}
\gamma^{0}\equiv\bar{\gamma}^{0} \ \ , \ \ \gamma^{j}\equiv i\bar{\gamma}^{j} \ \ ; \ \ j=1,2,\cdots , d	
\end{equation}
for which the following relations hold
\begin{equation}
\left\lbrace \bar{\gamma}^{\mu},\bar{\gamma}^{\nu} \right\rbrace =2\delta^{\mu\nu} \ , \ \left(\bar{\gamma}^{\mu} \right)^{\dag}=\bar{\gamma}^{\mu} \ , \ \text{Tr} \left(\bar{\gamma}^{\mu} \right)=0
\end{equation}
Here after, we consider the case $d=1$, for which the gamma matrices are $ 2\times 2 $. The generalization of the formalism to higher dimension is straightforward. For this case, one can easily show that the $\bar{\gamma}$ matrices have to be in the following form (see Eq.~\ref{Eq:gammaMatrice} Appendix):
\begin{equation}
\begin{split}
\bar{\gamma}^{\mu}&=\left(\begin{matrix}
c_{\mu} & a_{\mu}-ib_{\mu} \\
a_{\mu}+ib_{\mu} & -c_{\mu}
\end{matrix} \right) \\
&=a_{\mu}\sigma_x+b_{\mu}\sigma_y+c_{\mu}\sigma_z
\end{split}
\end{equation}
where $\mu=0,1$, $(a_{\mu},b_{\mu},c_{\mu}) \in\Re $, and $\sigma_x=\left(\begin{matrix}
0 & 1\\
1 & 0
\end{matrix} \right) $, $\sigma_y=i\left(\begin{matrix}
0 & -1\\
1 & 0
\end{matrix} \right) $, $\sigma_z=\left(\begin{matrix}
1 & 0\\
0 & -1
\end{matrix} \right) $ are Pauli matrices. For later convenience, let us define $\bar{\gamma}^2$ with the same definition as above, so that $\mu=0,1,2$ in the above equation. Using the anti-commutation relation of $\bar{\gamma}$ matrices, one can show that the general relation $a_{\mu}a_{\nu}+b_{\mu}b_{\nu}+c_{\mu}c_{\nu}=\delta_{\mu\nu}$ holds. One easily retrieves the standard representation (SR) limit by setting $c_{0}=b_{1}=a_{2}=1$, and zero for the others. \\

By constructing the general Dirac equation using this general $\gamma$ matrices one can readily find the plane wave solutions to be (see Eq.~\ref{Eq:plane-wave})
\begin{equation}
D_{\text{IFS}}\psi(E,k)\equiv \left( \bar{\gamma}^{0}E-i\bar{\gamma}^{1}k-mI \right)\psi(E,k)=0,
\end{equation}
where $k$ is the momentum of particle, and the explicit form of $D_{\text{IFS}}$ is
\begin{widetext}
\begin{equation}
	D_{\text{IFS}}\equiv
	\begin{pmatrix}
	c_{0}E-ic_{1}k-m & (a_{0}-ib_{0})E-i(a_{1}-ib_{1})k \\
	(a_{0}+ib_{0})E-i(a_{1}+ib_{1})k & -c_{0}E+ic_{1}k-m
	\end{pmatrix}
	\label{Eq:Difs}
	\end{equation}
\end{widetext}
By setting the determinant of $ D_{\text{IFS}}$ to zero, we recover the dispersion relation $ E^{2}=k^{2}+m^{2}$ ($E=\pm E_0$, where $E_0=\sqrt{k^2+m^2}$) as expected. The eigenfunctions are then given by 
\begin{equation}\label{Eq:eigenstates}
\psi_{\text{IFS}}^{+E_0}=\zeta
\begin{pmatrix}
1 \\
f^+
\end{pmatrix},\ \ \psi_{\text{IFS}}^{-E_0}=\zeta
\begin{pmatrix}
f^- \\
1
\end{pmatrix}
\end{equation}
where $f^{\pm}=\dfrac{i(b_{0}E_{0}-a_{1}k)\pm(a_{0}E_{0}+b_{1}k)}{c_{0}E_{0}+m\mp ic_{1}k}$, and $\zeta^2 = \dfrac{(c_{0}E_{0}+m)^{2}+c_{1}^{2}k^{2}}{2E_{0}(E_{0}-c_{2}k+c_{0}m)}$. This is a compact form of Eq.~\ref{Eq:eigenfunctions}, and gives the correct solution for the standard representation (SR), i.e. Eq.~\ref{Eq:standardSol} as the SR limit is taken. The other approach to get the above result is going to the moving reference (in which the particle is at rest, i.e. $k=0$), see Eqs.~\ref{Eq:restRef1} and~\ref{Eq:restRef2}, which leads consistently to a same result as Eq.~\ref{Eq:eigenstates} after the appropriate boost.\\
In the moving reference the eigenstates have to be simultaneously the eigenstates of the spin operator $S_z$, through which its shape can be found in this general representation. By requiring that $ S_{z}\psi_{\pm}=\pm\frac{1}{2}\psi_{\pm} $, it is not hard to find out that $	S_z=\dfrac{1}{2}\gamma^{0}=\dfrac{1}{2}\bar{\gamma}^{0} $. By going to the SR limit, one easily finds that $ S_{y} $ has no chance but following the relation $ S_{y}=\frac{1}{2}\bar{\gamma}^{1} $. Then using the fundamental commutation relation $ \left[S_{i},S_{j}\right]=i\epsilon_{ijk}S_{k} $, we can find $ S_{x} $ as Eq.~\ref{Eq:Sx}, which casts to
\begin{equation}
S_{x}\equiv \frac{1}{2}\bar{\gamma}^{2}=\frac{1}{2}
\begin{pmatrix}
c_{2} & a_{2}-ib_{2} \\ a_{2}+ib_{2} & -c_{2}
\end{pmatrix};
\end{equation}
with the following definitions
\begin{equation}
\begin{split}
\begin{cases}
a_{2}=c_{0}b_{1}-b_{0}c_{1} \\ b_{2}=a_{0}c_{1}-c_{0}a_{1}  \\ 
c_{2}=b_{0}a_{1}-a_{0}b_{1}
\end{cases},
\end{split}
\label{Eq:abc1}
\end{equation}
using of which one can easily show that
\begin{equation}
\begin{cases}
i\bar{\gamma}^{0}\bar{\gamma}^{1}\bar{\gamma}^{2}
=-i\gamma^{0}\gamma^{1}\gamma^{2}=I\\
\bar{\gamma}^{\mu}\bar{\gamma}^{\nu}=-i \epsilon^{\mu\nu\theta}\bar{\gamma}^{\theta}.
\end{cases} 
\end{equation}
where $\epsilon^{\mu\nu\theta}$ is totally antisymmetric symbol, and $\mu,\nu,\theta=0,1,2$. It should be noted that it is Hermitian and traceless, and satisfies the following relations
\begin{equation}
\left\lbrace S_{x},\bar{\gamma}^{0}\right\rbrace = \left\lbrace S_{x},\bar{\gamma}^{1}\right\rbrace =0\ ,\ 
S_{x}^{2}=\dfrac{1}{4}I
\end{equation}
These all can be easily generalized to $2+1$-dimensional space-time, using the same $\bar{\gamma}$s. \\

The other important question is concerning the spin representation of the particles. Based on the above-mentioned generalizations, we find that the general form of the spin operator is $S_{\mu}=\dfrac{1}{2}\bar{\gamma}^{\mu}$ with the following eigenstates
\begin{equation}
\left|S_{\mu}\pm\right\rangle =\dfrac{1}{\sqrt{2(1\pm c_{\mu})}}
\begin{pmatrix}
1\pm c_{\mu} \\ a_{\mu}+ib_{\mu} ,
\end{pmatrix}
\end{equation}
where $\mu=0$, $\mu=1$, and $\mu=2$ represent $S_z$, $S_y$ and $S_x$ respectively. This also helps to find the helicity operator for the rest frame ($k\ne 0$), i.e. $(h=\textbf{S}\cdot\textbf{p}/\mid\textbf{p}\mid)$, which is $h=S_{z}$ when the particle moves in the $z$ direction. Consequently, the right-hand and the left-hand side wave functions are $+$ and $-$ eigenstates of $S_{z}$ respectively.\\

One can easily prove that three independent matrices at most can be constructed for the case $d=1$, i.e. two dimensional $\gamma$ matrices.\\

Before finishing this section, let us summarize the relationships between the elements
\begin{equation}
\begin{split}
a_{\mu}a_{\nu}+b_{\mu}b_{\nu}+c_{\mu}c_{\nu}=\delta_{\mu\nu} \\ 
a_{\mu}a_{\mu}=b_{\mu}b_{\mu}=c_{\mu}c_{\mu}=1 \\
a_{\mu}b_{\mu}=a_{\mu}c_{\mu}=b_{\mu}c_{\mu}=0
\end{split}
\label{Eq:GeneralRelations1}
\end{equation}
which is eqivalent to
\begin{equation}
\begin{split}
&a_{\mu}=-b_{\nu}c_{\theta}\epsilon_{\mu\nu\theta} ,\\
&b_{\mu}=-c_{\nu}a_{\theta}\epsilon_{\mu\nu\theta}  ,\\
&c_{\mu}=-a_{\nu}b_{\theta}\epsilon_{\mu\nu\theta}
\end{split}
\label{Eq:GeneralRelations0}
\end{equation}
where $(\mu , \nu , \theta =0,1,2)$, and Einstein summation rule was used. \\
In the next section we re-shape the above equations in a single clean form, which is the main achievement of the present paper.

\subsection{Boost of IFSs}
A crucial question for any fermion that is governed by the Dirac equation is its behavior under the boost. Let us denote the space-time Lorentz transformation as $x'_{\mu}=\Lambda_{\mu\nu} x_{\nu}$, then the wave functions transform as $\psi'(x')=S(\Lambda)\psi(x)$, where $S(\Lambda)$ is a representation of the Lorentz transformation. Here the prime means inertial system $ O' $ that moves with velocity $ v=\tanh\theta\equiv\beta $ relative to the system $ O $. Therefore one can easily verify that $\Lambda_{0}^{0}=\Lambda_{1}^{1}=\cosh\theta$ and $ \Lambda_{0}^{1}=\Lambda_{1}^{0}=\sinh\theta$. In the 1+1-dimensional system we have just one boost direction, so that
\begin{equation}
\frac{E}{m}=\frac{k}{m\beta}=\frac{\sinh\theta}{\beta}=\cosh\theta\equiv\varGamma
\end{equation}
so that $\cosh\frac{\theta}{2}=\sqrt{\frac{\varGamma +1}{2}}=\sqrt{\frac{E_{0}+m}{2m}}$ and $\sinh\frac{\theta}{2}=\sqrt{\frac{\varGamma -1}{2}}=\sqrt{\frac{E_{0}-m}{2m}}$. In analogy with the boost of standard fermions, we examine the representation
\begin{equation}
\begin{split}
S&=\exp(\gamma^{0}\gamma^{1}\theta/2)=\exp(-i\bar{\gamma}^{0}\bar{\gamma}^{1}\theta/2)=\exp(-\bar{\gamma}^{2}\theta/2) \\
&=I\cosh\theta/2
+\bar{\gamma}^{2}\sinh\theta/2=\sqrt{\frac{E_{0}+m}{2m}}+\bar{\gamma}^{2}\sqrt{\frac{E_{0}-m}{2m}}
\end{split}
\end{equation}
which gives us the final result for the boost of the IFSs
\begin{equation}
S_{\text{IFS}}=
\begin{pmatrix}
\sqrt{\frac{E_{0}+m}{2m}}+c_{2}\sqrt{\frac{E_{0}-m}{2m}} & 
(a_{2}-ib_{2})\sqrt{\frac{E_{0}-m}{2m}} \\
(a_{2}+ib_{2})\sqrt{\frac{E_{0}-m}{2m}} & 
\sqrt{\frac{E_{0}+m}{2m}}-c_{2}\sqrt{\frac{E_{0}-m}{2m}}
\end{pmatrix}
\end{equation}
The general Dirac equation can be obtained using the above formula for the boost of IFS, see Appendix B for the details. To see if this formulation works, let us boost the solution in the rest reference ($\psi_{0\text{IFS}}$), for which we use Eq.~\ref{Eq:eigenstates}. The result is abbreviated as follows (see Eq.~\ref{Eq:restRef2})
\begin{equation}
\psi_{\text{IFS}}^{\pm m}(k=0)=\dfrac{1}{ \sqrt{2(1+ c_{0})}}
\begin{pmatrix}
\psi_1^{\pm} \\
\psi_2^{\pm}
\end{pmatrix}.
\end{equation}
where $\psi_1^+=\psi_2^-=1+c_0$, $\psi_2^+=a_{0}+ib_{0}$, and $\psi_1^-=-a_0+ib_0$. It should be taken into account that  $\psi_{\text{IFS}}^{\pm}(k)=S(\Lambda)\psi_{\text{IFS}}^{\pm m}(k=0)$ which is exactly the Eq.~\ref{Eq:eigenstates}. Now let us find a matrix which satisfies $D S_{\text{IFS}}\psi_{\text{IFS}}^{+ m}(k=0)=0$ which is the Dirac equation. To this end, we notice that 
\begin{equation}
S_{\text{IFS}}\psi_{\text{IFS}}^{+ m}(k=0)=\varrho \begin{pmatrix}
c_{0}E_{0}+m-ic_{1}k \\
(a_{0}+ib_{0})E_{0}-i(a_{1}+ib_{1}k)
\end{pmatrix}
\end{equation}
where
\begin{equation}
\varrho= \dfrac{\left[(1+c_{0})(E_{0}+m)-(c_{2}-ic_{1})k \right]\sqrt{E_{0}} }{\sqrt{2m(1+c_{0})(E_{0}+m)(E_{0}+c_{0}m-c_{2}k)}} .
\end{equation}
Then, by requiring that $\det D=E^{2}-k^{2}-m^{2}$ one readily finds $D=D_{\text{IFS}}$. This shows that one can reach to the wave function in general frame by a boost from the rest frame.

\section{$SO(3)$ symmetry in the representation of $\gamma$ matrices}\label{SEC:SO3}

In this section we aim to find the structure of the parameters that were obtained in the previous section, i.e. the relation between $a_{\mu}, b_{\mu}$ and $c_{\mu}$, $\mu=0,1,2$. To this end, let us put the parameters into a $3\times 3$ matrix $O$ as follows. 
\begin{equation}
O\equiv 
\begin{pmatrix}
a_{2} & a_{1} & a_{0} \\
b_{2} & b_{1} & b_{0} \\
c_{2} & c_{1} & c_{0}
\end{pmatrix}
\end{equation}
Note that $O_{S}=I$. At the first glance, it may seem ad hoc, but as will become clear soon, it helps much to view the transformation between IFS (representations of the Dirac equation) as a matrix operation. The interesting fact is that the conditions depicted in Eq.~\ref{Eq:GeneralRelations} can actually be written in the form 
\begin{equation}
OO^{T} =O^{T}O=I,
\end{equation} 
i.e. the matrix $O$ is orthogonal and reversible. As a result, the matrix $O$ is a member of $SO(3)$ group, so that various IFSs can be reached via rotation in this space. Let us show a rotation matrix with the angle $\varphi$ around the unit vector $
\hat{n}=n_{x}\hat{i}+n_{y}\hat{j}+n_{z}\hat{k}
$ as follows:
\begin{widetext}
\begin{equation}\label{Eq:transformation}
R_{n}(\varphi)=e^{-i\textbf{J}\cdotp \hat{n}\varphi}
=
\begin{pmatrix}
\cos\varphi + n_{x}^{2}(1-\cos\varphi)\ \  & 
n_{x}n_{y}(1-\cos\varphi)-n_{z}\sin\varphi\ \  & 
n_{x}n_{z}(1-\cos\varphi)+n_{y}\sin\varphi \\
n_{y}n_{x}(1-\cos\varphi)+n_{z}\sin\varphi\ \  & 
\cos\varphi + n_{y}^{2}(1-\cos\varphi)\ \  & 
n_{y}n_{z}(1-\cos\varphi)-n_{x}\sin\varphi  \\
n_{z}n_{x}(1-\cos\varphi)-n_{y}\sin\varphi\ \  & 
n_{z}n_{y}(1-\cos\varphi)+n_{x}\sin\varphi\ \  & 
\cos\varphi + n_{z}^{2}(1-\cos\varphi)
\end{pmatrix}
\end{equation}
\end{widetext}
so that $\textbf{J}\cdotp \hat{n}=i\dfrac{d R_{n}(\varphi)}{d\varphi}|_{\varphi=0}$. By matching elements of $O$ matrix with $R_{n}(\varphi)$ we obtain
\begin{align}
2\cos\varphi=a_{2}+b_{1}+c_{0}-1 \ \ \ \ \ \ \nonumber
\end{align}
in such a way that if $a_{2}=b_{1}=c_{0}=1$ then $\varphi=0 $, and if the other parameters are set to zero, then $R_{n}(\varphi)=I$ as expected. In general
\begin{equation}
\begin{cases}
2n_{z}\sin(\varphi)=b_{2}-a_{1} \\ 
2n_{y}\sin(\varphi)=a_{0}-c_{2} \\
2n_{x}\sin(\varphi)=c_{1}-b_{0}
\end{cases}
\end{equation}
and also
\begin{equation}
\begin{cases}
n_{x}n_{y}=\dfrac{b_{2}+a_{1}}{3-a_{2}-b_{1}-c_{0}} \\ 
n_{x}n_{z}=\dfrac{a_{0}+c_{2}}{3-a_{2}-b_{1}-c_{0}} \\
n_{y}n_{z}=\dfrac{c_{1}+b_{0}}{3-a_{2}-b_{1}-c_{0}}.
\end{cases}
\end{equation}
The above equations give us the full correspondence between the space of representation of the Dirac equation (shown by $O$ matrices) and the general representation of $SO(3)$ group. Using the correspondence between $SO(3)$ and SU(2) groups, one can associate the representation of the IFSs with SU(2) group. We make this correspondence using the $(S_{\mu})$ that we found in the previous section as the generators of $SU(2)$. More precisely, let us define
\begin{equation}
\begin{split}
U&=e^{-i\textbf{S}\cdot\hat{n}\varphi}\\
&=\begin{pmatrix}
\cos\varphi/2 - i n_{\mu}c_{\mu}\sin\varphi/2 & 
-n_{\mu}(b_{\mu}+i a_{\mu})\sin\varphi/2 \\
n_{\mu}(b_{\mu}-i a_{\mu})\sin\varphi/2 & 
\cos\varphi/2 + i n_{\mu}c_{\mu}\sin\varphi/2
\end{pmatrix}
\end{split}
\end{equation}
where $\det(U)=1$. An example is $ n_{\mu}=a_{\mu}$ for which $U=\begin{pmatrix}
\cos\varphi/2 & 
-i \sin\varphi/2 \\
-i \sin\varphi/2 & 
\cos\varphi/2
\end{pmatrix}$.
Generally if we define 
\begin{equation}
M=
\begin{pmatrix}
c_{\nu}x_{\nu} & 
(a_{\nu}-i b_{\nu})x_{\nu} \\
(a_{\nu}+i b_{\nu})x_{\nu} & 
-c_{\nu}x_{\nu}
\end{pmatrix}
\end{equation}
then the transformed matrix $M'=UMU^{\dag}=\begin{pmatrix}
c_{\nu}x'_{\nu} & 
(a_{\nu}-i b_{\nu})x'_{\nu} \\
(a_{\nu}+i b_{\nu})x'_{\nu} & 
-c_{\nu}x'_{\nu}
\end{pmatrix}$ is such that $\begin{pmatrix}
x' & y'& z'
\end{pmatrix}=
\begin{pmatrix}
x & y & z
\end{pmatrix}R_{n}^{\top}(\varphi)$.

\section{sub-representations of IFS}~\label{SEC:subr}
By ``sub-representation'', we mean \textit{restricted} $\gamma$ representations. For instance, let us consider the Majorana representation for which fermions and antifermions are the same, limiting strongly the range of the entities of $\gamma$ matrices. Fermions ($\psi$) and antifermions ($\psi_c$ obtained by charge conjugation) satisfy the Dirac equation in the presence of electromagnetic field ($A_{\mu}$)
\begin{equation}
\begin{cases}
\left[\gamma^{\mu}(p_{\mu}-eA_{\mu})-m \right]\psi=0  \\ 
\left[\gamma^{\mu}(p_{\mu}+eA_{\mu})-m \right]\psi_c=0
\end{cases}
\end{equation}
If there is a transformation $U$ such that $U(\gamma^{\mu})^{*}U^{-1}=-\gamma^{\mu}$, then one can show by inspection that $U\psi^{*}$ is a solution of the second equation, giving us no chance but $\psi_c=e^{i\alpha}U\psi^*$ where $\alpha$ is an arbitrary phase. These fermions are Majorana, in which, for the simple case $U=I$ (identity matrix), the wave function of fermions and antifermions are the same. Without loss of generality, we set $U=I$ in this paper (in other cases we always can transform $\gamma$ so that it applies). Let us call the $\gamma$ matrices that satisfy this condition constitute the \textit{general Majorana representation}, which are
\begin{equation}\label{Eq:MajoranaGamma}
\gamma^{0}_{\text{M}\pm}=\pm i
\begin{pmatrix}
0 & -1 \\
1 & 0
\end{pmatrix} \ , \ 
\gamma^{1}_{\text{M}\pm}= i
\begin{pmatrix}
c_{1} & \pm\sqrt{1-c_{1}^{2}} \\
\pm\sqrt{1-c_{1}^{2}} & -c_{1}
\end{pmatrix}.
\end{equation}
were ``M'' stands for ``Majorana'', and the ``$\pm$'' for $\gamma^0_{\text{M}}$ is independent of ``$\pm$'' in $\gamma^1_{\text{M}}$, so that we have four possibilities for selecting $\gamma$s, i.e. $(\gamma_{\text{M}+}^0,\gamma_{\text{M}+}^1)$, $(\gamma_{\text{M}+}^0,\gamma_{\text{M}-}^1)$, etc. Then, for example for the $(+,+)$ case, using these matrices the eigenstates are calculate to be
\begin{equation}\label{Eq:majorana}
\psi_{\eta,\eta'}^{\text{M}}(E,k)=A_0
\begin{pmatrix}
m-ic_{1}k \\
i\left(\eta E-\eta'\sqrt{1-c_{1}^{2}}k\right) 
\end{pmatrix}
\end{equation}
where $A_0=\dfrac{1}{\sqrt{2E(E-\eta\eta'\sqrt{1-c_{1}^{2}}k)}}$ and $\eta , \eta'=\pm$ refer to the signs of $\gamma^0$ and $\gamma^1$ respectively.

\begin{figure*}
	\begin{subfigure}{0.3\textwidth}\includegraphics[width=\textwidth]{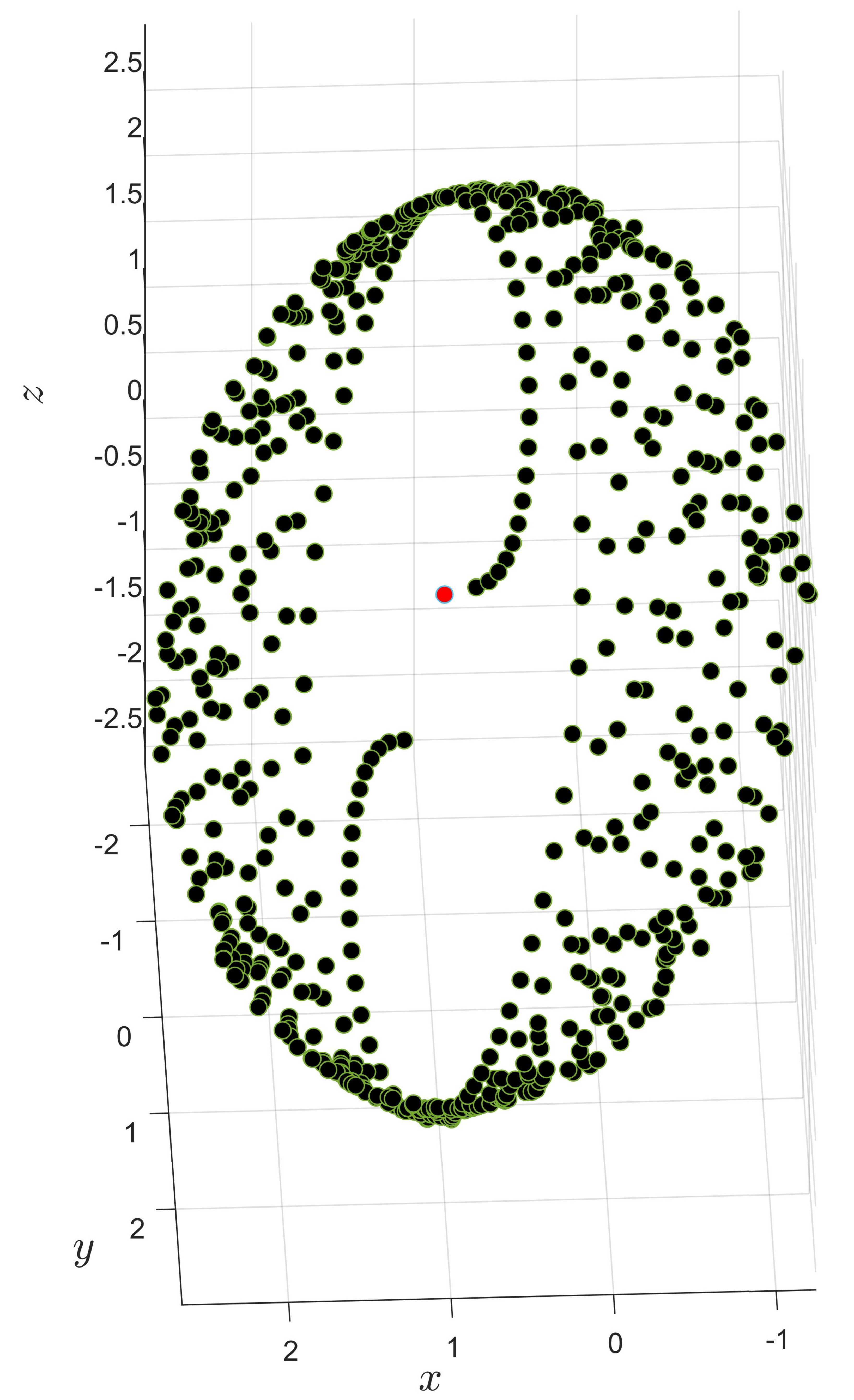}
		\caption{}
		\label{fig:U(1)a}
	\end{subfigure}
	\begin{subfigure}{0.5\textwidth}\includegraphics[width=\textwidth]{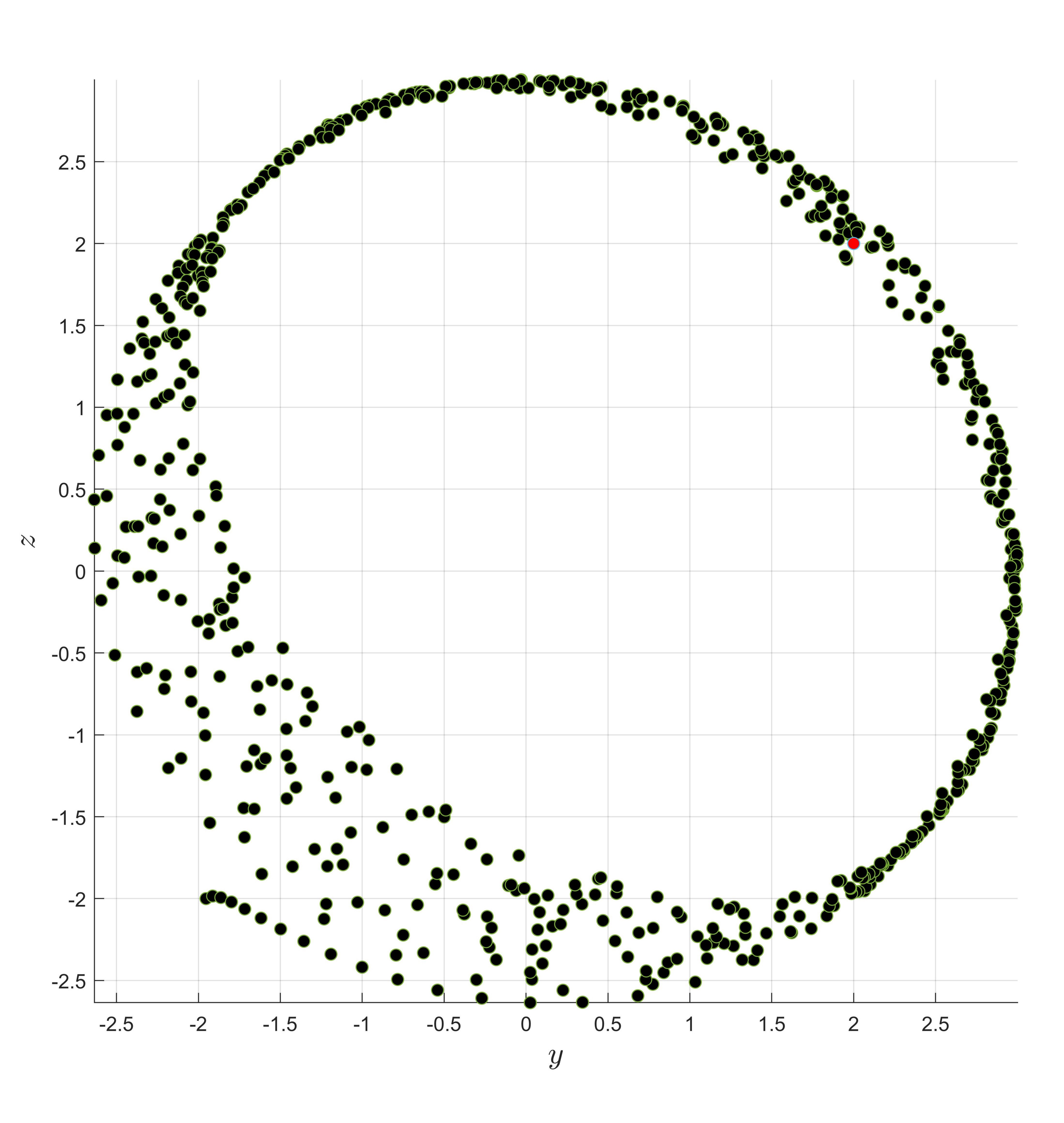}
		\caption{}
		\label{fig:U(1)b}
	\end{subfigure}
	\caption{Three-dimensional representation of the rotation of Eq.~\ref{Eq:one-dimensional} from two views in terms of one parameter $c_1$ for $\eta=\eta'=1$. The start point $(x_0,y_0,z_0)$ is represented by red circle. }
	\label{fig:U(1)}
\end{figure*}

For Weyl-Majorana (WM) fermions, one sets $m=0$, for which the $\gamma$ matrices have to satisfy ${\gamma^{\mu}}^*=\pm\gamma^{\mu}$ (note that ``$-$'' sign is also permissible), and $\Psi_{\text{WM}}^*\propto \Psi_{\text{WM}}$. If $\left(\gamma^{\mu} \right)^{*}=-\gamma^{\mu}$ we get previous representation just above, but for the opposite case we have
\begin{equation}
\begin{split}
\gamma^{0}_{\text{WM}\pm}&=
\begin{pmatrix}
c_{0} & \pm\sqrt{1-c_{0}^{2}} \\
\pm\sqrt{1-c_{0}^{2}} & -c_{0}
\end{pmatrix} \ ,\ 
\gamma^{1}_{\text{WM}\pm}=
\begin{pmatrix}
0 & \pm 1 \\
\mp 1 & 0
\end{pmatrix}
\end{split}
\end{equation}
for which the eigenstates are
\begin{equation}
\psi^{\text{WM}}_{\eta,\eta'}=\dfrac{1}{\sqrt{2(1-\eta\eta'\sqrt{1-c_{0}^{2}})}}
\begin{pmatrix}
c_{0} \\
\eta\sqrt{1-c_{0}^{2}}- \eta'
\end{pmatrix}
\end{equation}
Note that in the above equation $E$ and $k$ cancel out, so that its form is simpler than Eq.~\ref{Eq:majorana}. \\

It is worth mentioning that we can reach the Majorana representation starting from the Dirac equation by a rotation. To be more precise, if we ``rotate'' the $\gamma$ matrices in the Dirac representation about  $\hat{n}$ by the angle $\varphi$, which are
\begin{equation}\label{Eq:one-dimensional}
\begin{split}
&n_{x}=\pm\eta\sqrt{\frac{1-\eta' c_{1}}{3+\eta' c_{1}}},
n_{y}=\pm\sqrt{\frac{1+\eta' c_{1}}{3+\eta' c_{1}}},
n_{z}=\pm\eta'\sqrt{\frac{1+\eta' c_{1}}{3+\eta' c_{1}}} \\
&\cos\varphi=\frac{-1}{2}(1+\eta'c_{1}),
\sin\varphi=\frac{\mp\eta\eta'}{2}\sqrt{3-2\eta'c_{1}-c_{1}^{2}}
\end{split}
\end{equation}
then we reach the $\gamma$ matrices given in Eq.~\ref{Eq:MajoranaGamma}. We notice here that all quantities depend on a single parameter, i.e. $c_1$, showing that the transformation is isomorphism to $U(1)$ group, which is a subgroup of $SU(2)$, which itself is homomorphism to $SO(3)$. In the Fig.~\ref{fig:U(1)} we show a three-dimensional representation of this transformation. In this figure a starting point $(x_0,y_0,z_0)$ (which is represented by a red bold circle) under the effect of Eq.~\ref{Eq:one-dimensional} and Eq.~\ref{Eq:transformation}. This can also be done easily for WM fermions.

\section{Transformation of Dirac to generalized particles}\label{SEC:Trans}

In this section we find the general transformations $T$ using of which IFSs are obtained from standard (Dirac) representation. For the definition of $T$ see Eq.~\ref{Eq:transformation-gamma}. Using the calculations presented in Appendix C one can show that
\begin{equation}
T=\varrho'\begin{pmatrix}
1+q_{1}q_{2}e^{i\alpha} & -(q_{1}-q_{2}e^{-i\alpha}) \\ q_{1}-q_{2}e^{i\alpha} & 1+q_{1}q_{2}e^{-i\alpha}
\end{pmatrix}
\end{equation} 
where $\varrho'\equiv \frac{1}{2}E_0^{-1}\sqrt{\left(E_0+m \right) \left(E_0+c_0m+c_2k \right) }$ and $\alpha=\tan^{-1}\frac{b_0m+b_2k}{a_0m+a_2k}$, and
\begin{equation}
q_1=\sqrt{\dfrac{E_{0}-m}{E_{0}+m}} \ , \ q_{2}=\sqrt{\dfrac{E_{0}-c_{0}m-c_{2}k}{E_{0}+c_{0}m+c_{2}k}}
\end{equation}
One can easily show that $TT^{\dag}=T^{\dag}T=I$ and $\det T=1$, showing that they are unitary transformations. These matrices are also represented by $T= e^{-i\sigma\cdot\hat{n}\varphi/2}$, were $\varphi$ is a real parameter, rotating $\Psi_{_{S}}$ to $\Psi_{_{IFS}}$, i.e. $T\psi_{\text{S}}=\psi_{\text{IFS}}$. Using this notation, one can easily find the rotation parameters, represented by $\hat{n}=(n_x,n_y,n_z)$, satisfying the following identities
\begin{widetext}
\begin{equation}
\begin{split}
&n_{x}\sin\varphi/2=\frac{1}{2}\sin\alpha\sqrt{\left(1+\frac{m}{E} \right) \left(1-c_{0}\frac{m}{E}-c_{2}\frac{k}{E} \right) }      \\
&n_{y}\sin\varphi/2=\frac{1}{2}\left[\sqrt{\left(1-\frac{m}{E} \right) \left(1+c_{0}\frac{m}{E}+c_{2}\frac{k}{E} \right) }-\cos\alpha\sqrt{\left(1+\frac{m}{E} \right) \left(1-c_{0}\frac{m}{E}-c_{2}\frac{k}{E} \right) } \right]       \\
&n_{z}\sin\varphi/2=-\frac{1}{2}\sin\alpha\sqrt{\left(1-\frac{m}{E} \right) \left(1-c_{0}\frac{m}{E}-c_{2}\frac{k}{E} \right) }       \\
&\cos\varphi/2=\frac{1}{2}\left[\sqrt{\left(1+\frac{m}{E} \right) \left(1+c_{0}\frac{m}{E}+c_{2}\frac{k}{E} \right) }+\cos\alpha\sqrt{\left(1-\frac{m}{E} \right) \left(1-c_{0}\frac{m}{E}-c_{2}\frac{k}{E} \right) } \right] 
\end{split}
\end{equation}
\end{widetext}
from which one can show $n_{x}^{2}+n_{y}^{2}+n_{z}^{2}=1$. As an example, let us consider the transformation $T_{\text{S-M}}$ that converts a standard Dirac particle to a Majorana particle
\begin{equation}
T_{\text{S-M}}=\varrho''
\begin{pmatrix}
1+q'_{1}q'_{2}e^{i\alpha'} & -(q'_{1}-q'_{2}e^{-i\alpha'}) \\ q'_{1}-q'_{2}e^{i\alpha'} & 1+q'_{1}q'_{2}e^{-i\alpha'}
\end{pmatrix}
\end{equation}
were $\varrho''=\frac{1}{2}E_0^{-1}\sqrt{(E_{0}+m)(E_{0}+c_{2}k)}$, $\alpha'=\tan^{-1}(-m/c_{1})$, $q'_{1}=q_{1}$, and $q'_{2}=\sqrt{\dfrac{E_{0}-c_{2}k}{E_{0}+c_{2}k}}$.

\section{Conclusion}
In this paper we considered a general form for the $\gamma$ matrices. Motivated by the fact that the resulting fermions are ``intermediate'' in the sense of normal representation (i.e. standard representation, Weyl representation, etc.) we call them the ``intermediate fermion species'' (IFS). Many properties of the IFS were calculated and explored, like the eigenvalue problem, boost and rotation, and transformation between species. We observed that the latter (the transformation between species) corresponds to $SO(3)$ rotations in the space of the parameters of the problem (the entities of the $\gamma$ matrices). Therefore any arbitrary representation of spin $\frac{1}{2}$ fermions is obtained by a $SO(3)$ rotation in the parameters of the $\gamma$ matrices. Based on this, we calculated the sub-representations which admits the Majorana fermions. Importantly, we clearly established that any IFS can be obtained from the Dirac spinors by a SU(2) similarity transformation. \\

It is worth mentioning that this transformation does not change the transport properties of particles. For instance, we measured the transport parameters of the Klein tunneling, and noticed that none of these parameters (reflection and transmission coefficients) change under the mentioned $SO(3)$ transformation in normal incidence. This motivated us to call it ``the symmetry'' of the Dirac equation.\\

According to the Noether theorem this symmetries leads to some conservations between IFS particles. In our future research we intend to concentrate on this topic, and also finding the other aspects of this transformation, such as Andreev reflection, to see if we can design an experiment which distinguish between these particles.

\appendix

\section{The properties of $\bar{\gamma}$ matrices}
For interested readers a detailed description of first subsection in section two is presented in this Appendix.
 In $1+1$, $\bar{\gamma}$ matrices should be of the following form
\begin{equation}
\bar{\gamma}^{\mu}=
\begin{pmatrix}
c_{\mu} & a_{\mu}-ib_{\mu} \\
a_{\mu}
+ib_{\mu} & -c_{\mu}
\end{pmatrix}
\ ; \ a_{\mu},b_{\mu},c_{\mu} \in\Re 
\label{Eq:gammaMatrice}
\end{equation}
for which $\mu,\nu=0,1$. Using the anticommutation relation of $\bar{\gamma}$ matrices (Eq. 3), one can generally show that 
\begin{equation}
a_{\mu}a_{\nu}+b_{\mu}b_{\nu}+c_{\mu}c_{\nu}=\delta_{\mu\nu}
\end{equation}
In the 1+1 dimension,
according to the what had been said, we only need $ \gamma^{0} $ and $ \gamma^{1} $
\begin{equation}
\begin{split}
&\left( \gamma^{0}E-\gamma^{1}k-mI \right)\psi(E,k) \ e^{i(kx-Et)}= \\
&\left( \bar{\gamma}^{0}E-i\bar{\gamma}^{1}k-mI \right)\psi(E,k) \ e^{i(kx-Et)}=0
\label{Eq:plane-wave}
\end{split}
\end{equation}
therefore, the non-differential form of intermediate fermionic species (IFS) Dirac equation will be Eq. 8.
We have non-trivial answer, if $ E^{2}=k^{2}+m^{2}\Rightarrow E=\pm E_{0}=\pm\sqrt{k^{2}+m^{2}} $ , which we also expect it before. General shape of free particle’s wave function in $1+1$ dimensions is:

	\begin{equation}\label{Eq:eigenfunctions}
	\begin{cases}
\psi_{\text{IFS}}^{+E_0}=\zeta
	\begin{pmatrix}
	1 \\
	\dfrac{i(b_{0}E_{0}-a_{1}k)+(a_{0}E_{0}+b_{1}k)}{c_{0}E_{0}+m-ic_{1}k}
	\end{pmatrix}
\\ \\
\psi_{\text{IFS}}^{-E_0}=\zeta
	\begin{pmatrix}
	\dfrac{i(b_{0}E_{0}-a_{1}k)-(a_{0}E_{0}+b_{1}k)}{c_{0}E_{0}+m+ic_{1}k} \\
	1
	\end{pmatrix}
	\end{cases}
\end{equation}
	where  $ \zeta =\sqrt{\dfrac{(c_{0}E_{0}+m)^{2}+c_{1}^{2}k^{2}}
		{2E_{0}(E_{0}-c_{2}k+c_{0}m)}}$.
For Standard Representation limit $(c_{0}=b_{1}=a_{2}=1 \ , \ etc=0)$ they Turns into:
\begin{equation}\label{Eq:standardSol}
\begin{cases}
\psi_{_{S+}}=\sqrt{\frac{E_{0}+m}{2E_{0}}}
\begin{pmatrix}
1 \\
\frac{+k}{E+m}
\end{pmatrix}
\\ \psi_{_{S-}}=\sqrt{\frac{E_{0}+m}{2E_{0}}}
\begin{pmatrix}
\frac{-k}{E+m} \\
1
\end{pmatrix} 
\end{cases}
\end{equation}
The non-differential Dirac equation in $k=0$ reduces as follows:
\begin{equation}\label{Eq:restRef1}
\begin{pmatrix}
c_{0}E-m & (a_{0}-ib_{0})E \\
(a_{0}+ib_{0})E & -c_{0}E-m
\end{pmatrix}
\psi_{_{0IFS}}
=0\Rightarrow E=\pm m 
\end{equation}
\begin{equation}\label{Eq:restRef2}
\Rightarrow
\begin{cases}
 E_{+}=+m\Rightarrow
\psi_{_{0IFS+}}=\sqrt{\dfrac{1+c_{0}}{2}}
\begin{pmatrix}
1 \\ \dfrac{a_{0}+ib_{0}}{1+c_{0}}
\end{pmatrix}
\\
 E_{-}=-m\Rightarrow
\psi_{_{0IFS-}}=\sqrt{\dfrac{1+c_{0}}{2}}
\begin{pmatrix}
-(\dfrac{a_{0}-ib_{0}}{1+c_{0}}) \\ 1
\end{pmatrix}.
\end{cases}
\end{equation}
We know the answers of Dirac equation in the case k=0, namely $ \psi_{\pm} $, must also be the eigenstates of the operator $ S_{z} $. So, we look for the matrix form of operator $ S_{z} $ such that $ S_{z}\psi_{\pm}=\pm\frac{1}{2}\psi_{\pm} $ are its eigenstates. Then $	S_{z}=\dfrac{1}{2}\gamma^{0}=\dfrac{1}{2}\bar{\gamma}^{0} $.
By respecting that $ \bar{\gamma}^{1} $ in the limit SR, becomes similar to $ S_{y} $, we can assume that $ S_{y}=\frac{1}{2}\bar{\gamma}^{1} $. Now, by using the fundamental commutation relation $ \left[S_{i},S_{j}\right]=i\in_{ijk}S_{k} $, we can also obtain $ S_{x} $:
\begin{widetext}
	\begin{equation}
	2S_{x}=
	\begin{pmatrix}
	(b_{0}a_{1}-a_{0}b_{1}) & (c_{0}b_{1}-b_{0}c_{1})-i(a_{0}c_{1}-c_{0}a_{1}) \\ (c_{0}b_{1}-b_{0}c_{1})+i(a_{0}c_{1}-c_{0}a_{1}) &-(b_{0}a_{1}-a_{0}b_{1})
	\end{pmatrix} .
	\label{Eq:Sx}
	\end{equation}
\end{widetext}
It is clear that the matrix $ S_{x} $ is Hermitian and traceless. By performing the respected calculations, we can see that:
\begin{equation}
\left\lbrace S_{x},\bar{\gamma}^{0}\right\rbrace = \left\lbrace S_{x},\bar{\gamma}^{1}\right\rbrace =0,
S_{x}\bar{\gamma}^{0}\neq 0 \neq S_{x}\bar{\gamma}^{1},
S_{x}^{2}=\dfrac{1}{4}I
\end{equation}
According to (4) condition, it can be say that $ 2S_{x} $ has the all conditions of a $ \bar{\gamma} $ (Eq. 10)

\section{Obtaining Standard Dirac equation by Lorentz operator}

In this Appendix we want to obtain the non-differential form of Dirac equation in Standard Representation by acting of Lorentz operator on wave function in rest framework i.e. $\psi_{_{S0+}}=
	\begin{pmatrix}
	1 \\
	0
	\end{pmatrix} ,
	\psi_{_{S0-}}=
	\begin{pmatrix}
	0 \\ 1
	\end{pmatrix}$. Lorentz operator in (1+1) dimension in Standard Representation is
\begin{equation}
S_{_{S}}=
\begin{pmatrix}
\sqrt{\frac{E_{0}+m}{2m}} & 
\sqrt{\frac{E_{0}-m}{2m}} \\
\sqrt{\frac{E_{0}-m}{2m}} & 
\sqrt{\frac{E_{0}+m}{2m}}
\end{pmatrix},
\end{equation}
then
\begin{equation}
S_{_{S}}\psi_{_{S0+}}=\sqrt{\frac{E_{0}}{m}}\psi_{_{S+}}\sim
\begin{pmatrix}
E_{0}+m \\
k
\end{pmatrix}.
\end{equation}
If we introduce Dirac equation with $D_{_{S}}$ then $D_{_{S}}S_{_{S}}\psi_{_{S0+}}=D_{_{S}}\psi_{_{S+}}=0$. By requiring that 
$det(D_{_{S}})=E_{0}^{2}-k^{2}-m^{2}=0$ one readily find
\begin{equation}
	 D_{_{S}}=
\begin{pmatrix}
E-m & -k \\
+k & -(E+m)
\end{pmatrix}\psi_{_{S}}=0
\end{equation}
that is exactly the Dirac equation known in (1+1).\\

\section{Transformation matrix of various spinors}
Standard Dirac Hamiltonian and Intermediate Fermionic Species Dirac Hamiltonian are
\begin{widetext}
\begin{equation}
	\begin{split}
	H_{_{S}}=
\begin{pmatrix}
	m & k \\
	k & -m
\end{pmatrix}\qquad , \qquad
	H_{_{IFS}}=
\begin{pmatrix}
c_{0}m+c_{2}k & (a_{2}-ib_{2})k+(a_{0}-ib_{0})m \\
(a_{2}+ib_{2})k+(a_{0}+ib_{0})m & -(c_{0}m+c_{2}k)
\end{pmatrix}
\end{split}.
\end{equation}
\end{widetext}
We know that eigenvalues of both Hamiltonians are the same $(\pm E_{0})$. Then
 in according to eigenstates of those, one can obtain nonsingular matrices $\Theta_{_{S}}$ and$\Theta_{_{IFS}}$ as follows
 \begin{equation}
\begin{split}
	\Theta_{_{S}}=
\begin{pmatrix}
	\sqrt{\frac{E_{0}+m}{2E_{0}}} & -\sqrt{\frac{E_{0}-m}{2E_{0}}} \\
	\sqrt{\frac{E_{0}-m}{2E_{0}}} & \sqrt{\frac{E_{0}+m}{2E_{0}}}
\end{pmatrix} \qquad \qquad\\	
\Theta_{_{IFS}}=
\begin{pmatrix}
	\sqrt{\frac{E_{0}+c_{0}m+c_{2}k}{2E_{0}}} & -e^{-i\alpha}\sqrt{\frac{E_{0}-c_{0}m-c_{2}k}{2E_{0}}} \\
	e^{i\alpha}\sqrt{\frac{E_{0}-c_{0}m-c_{2}k}{2E_{0}}} & \sqrt{\frac{E_{0}+c_{0}m+c_{2}k}{2E_{0}}}
\end{pmatrix} 	
	\end{split}
 \end{equation}
 where $\alpha=tan^{-1}\left( \dfrac{b_{0}m+b_{2}k}{a_{0}m+a_{2}k}\right)$, then
\begin{equation}
\Theta_{_{S}}^{\dag}H_{_{S}}\Theta_{_{S}}=\Theta_{_{IFS}}^{\dag}H_{_{IFS}}
\Theta_{_{IFS}}=
\begin{pmatrix}
E_{0} & 0 \\ 0 & -E_{0}
\end{pmatrix}.
\end{equation}
This suggest that $\Theta_{_{IFS}}\Theta_{_{S}}^{\dag}H_{_{S}}\Theta_{_{S}}\Theta_{_{IFS}}^{\dag}
=H_{_{IFS}}$. With definition $\Theta_{_{S}}\Theta_{_{IFS}}^{\dag}\equiv \vartheta$ one can see that $\vartheta\psi_{_{S}}=\psi_{_{IFS}}$, so that
\begin{equation}
\vartheta=\varrho'\begin{pmatrix}
1+q_{1}q_{2}e^{i\alpha} & -(q_{1}-q_{2}e^{-i\alpha}) \\ q_{1}-q_{2}e^{i\alpha} & 1+q_{1}q_{2}e^{-i\alpha}
\end{pmatrix}
\end{equation}
where
\begin{equation} 
\begin{split}
\varrho'=\dfrac{\sqrt{(E_{0}+m)(E_{0}+c_{0}m+c_{2}k)}}{2E_{0}} \quad \\ \alpha=tan^{-1}\left( \dfrac{b_{0}m+b_{2}k}{a_{0}m+a_{2}k}\right) \qquad
\\
q_{1}=\sqrt{\dfrac{E_{0}-m}{E_{0}+m}} \ ,\
 q_{2}=\sqrt{\dfrac{E_{0}-c_{0}m-c_{2}k}{E_{0}+c_{0}m+c_{2}k}}
\end{split}
\end{equation}

\bibliography{refs}

\end{document}